\documentclass[superscriptaddress,nofootinbib,preprint]{revtex4-1}

\usepackage{hyperref}
\usepackage{bbm}
\usepackage{amsfonts}
\usepackage{mathrsfs}

\usepackage{amsmath,amssymb}
\usepackage{epsfig,amsmath,amssymb,verbatim,mathrsfs,array,layout,textcomp,amssymb,latexsym, slashed,graphicx}

\usepackage{epsfig}
\usepackage{graphicx}               
\usepackage{url}
\usepackage{hyperref}
\usepackage{float}
\usepackage{color}
\usepackage{multirow}
\makeatletter
\def\simgt{\mathrel{\lower2.5pt\vbox{\lineskip=0pt\baselineskip=0pt
           \hbox{$>$}\hbox{$\sim$}}}}
\def\simlt{\mathrel{\lower2.5pt\vbox{\lineskip=0pt\baselineskip=0pt
           \hbox{$<$}\hbox{$\sim$}}}}
\makeatother

\def\mysection#1{{\bf #1.} }

\newcommand{\be}{\begin{equation}}
\newcommand{\ee}{\end{equation}}
\newcommand{\bea}{\begin{eqnarray}}
\newcommand{\eea}{\end{eqnarray}}
\newcommand{\beq}{\begin{eqnarray}}
\newcommand{\eeq}{\end{eqnarray}}

\def\lsim{\mathrel{\rlap{\lower4pt\hbox{\hskip1pt$\sim$}}
     \raise1pt\hbox{$<$}}}         
\def\gsim{\mathrel{\rlap{\lower4pt\hbox{\hskip1pt$\sim$}}
     \raise1pt\hbox{$>$}}}         

\begin{document}

\thispagestyle{empty}

\title{Modular Fluctuations from Shockwave Geometries}
\author{Erik Verlinde}
\affiliation{Institute of Physics, University of Amsterdam, Amsterdam, The Netherlands}
\author{Kathryn M. Zurek}
\affiliation{Walter Burke Institute for Theoretical Physics, \\
California Institute of Technology, Pasadena, CA USA}

\begin{abstract}
\noindent 
Modular fluctuations have previously been shown to obey an area law $\langle \Delta K^2 \rangle = \langle K \rangle = 
{A}/{4 G_N}$.  Furthermore, modular fluctuations generate fluctuations in the spacetime geometry of empty causal diamonds.  Here we demonstrate the physical origin of these fluctuations, showing that the modular area law, in $d-$dimensionsal Minkowski space, can be reproduced from shockwaves arising from vacuum fluctuations.  The size of the vacuum fluctuations is fixed by commutation relations in light-ray operators, of the same form postulated by 't Hooft in the context of black hole horizons. 
 \end{abstract}

\maketitle

\tableofcontents

\section{Introduction}
An important question in quantum gravity is to determine the size of the quantum fluctuations in the spacetime geometry.  Recent advances in understanding the (emergent) properties of gravity and spacetime in connection to quantum information theory offer a concrete path towards answering this question. A central role in these developments is played by the entanglement entropy associated with a finite region of spacetime bounded by a (Rindler) horizon. A related physical quantity that enters in these studies is the modular Hamiltonian. Using path integral techniques similar to those that were used in deriving the Ryu-Takayanagi formula, one can derive the fluctuations in the modular Hamiltonian. The goal of this paper is to relate these modular energy fluctuations to metric perturbations that take the form of gravitational shockwaves. 

Gravitational shockwaves have provided fundamental insight into the structure of quantum gravity.  Developed to study the effect of outgoing Hawking radiation on energetic particles traveling along the horizon of a black hole \cite{Dray:1984ha,Dray:1985yt,tHooft:1996rdg,tHooft:2018fxg}, the shockwave geometries have proved to provide insight where the quantum entanglement structure changes under the backreaction from infalling and outgoing matter, particularly in the context of black holes and the AdS/CFT correspondence \cite{Mertens:2017mtv,Lam:2018pvp}.  More recently, there have been efforts to extend this work beyond black holes and negatively curved space into flat space, including a thermodynamic description of gravitational shockwaves as small variations in the Bekenstein-Hawking area law  \cite{Liu:2021kay}.

In this paper we are interested in vacuum fluctuations in the spacetime geometry due to shockwaves resulting from vacuum energy fluctuations. In particular, quantum fluctuations in the energy momentum tensor, having longitudinal light cone components, $T_{uu},~T_{vv}$, generate longitudinal light cone metric fluctuations $h_{uu},~h_{vv}$ of the form   
\beq
h_{uu}(u,y)  = \ell_p^{d-2} \int \!d^{d-2}y' f(y,y') \,T_{uu}(u,y'),~~~ h_{vv}(u,y)  = \ell_p^{d-2} \int \!d^{d-2}y' f(y,y') \,T_{vv}(u,y')
\label{eq:shockwavemetric}
\eeq
where $\ell_p^{d-2} = 8 \pi G_N$.  Here $f(y,y')$ represents the Green function of the transversal Laplacian $\Delta_y$ and obeys
\beq
\label{Green}
	\Delta_y f(y,y') = \delta^{(d-2)}(y,y').
\eeq
The shockwave geometries lead to tiny shifts $\delta u(y)$ and $\delta v(y)$ in the longitudinal light coordinates $u$ and $v$. Applying the arguments of 't Hooft at a black hole horizon \cite{tHooft:1996rdg,tHooft:2018fxg} to a light sheet horizon, it can be shown that these shifts obey uncertainty relations of the form
\beq
\Delta \delta u(y)\, \Delta \delta v(y') = l^{d-2}_p f(y,y'),
\label{eq:uncertainty}
\eeq
where $\Delta$ here denotes the uncertainty on the quantum shifts $\delta u(y),~\delta v(y)$.

The purpose of this paper is to relate such shockwave geometries to the vacuum expectation value and fluctuations of the modular Hamiltonian $K$ associated with a spacetime region bounded by a Rindler horizon.  
The modular Hamiltonian $K$ is defined microscopically in terms of the density matrix $\rho$ obtained by tracing out the complement of the region via
\begin{equation}
\rho = \frac{e^{- K}}{Z}\qquad\mbox{with}
\qquad Z={\rm tr}\bigl( e^{- K}\bigr).
\end{equation}
One can normalize $K$ so that in the vacuum $Z=1$, so that its vacuum expectation value is precisely equal to the entanglement entropy $S=-{\rm tr}(\rho\log\rho)$, and one finds 
\begin{equation}
	S = \langle K\rangle = {Area\over 4 G}.
\end{equation}
Thus the quantum entanglement of a region bounded by a Rindler horizon is characterized by the entanglement entropy, which in magnitude is given by $\langle K \rangle$. This result can be derived from the gravitational replica trick, following the same steps as in \cite{Lewkowycz:2013nqa}. These same methods can be used to compute the fluctuations in $K$.  It has by now been firmly established, both in AdS \cite{perl,Nakaguchi:2016zqi,deBoer:2018mzv,VZ2} and in flat space \cite{BZ,Gukov:2022oed}, that these modular  energy fluctuations $\Delta K$ are also determined by the area of the horizon via the relation
\begin{equation}
\Bigl\langle \Delta K^2\Bigr\rangle = {Area\over 4 G}	\qquad \mbox{with}\qquad \Delta K = K-\langle K\rangle. 
\label{eq:DeltaK}
\end{equation}
Here $\Delta K$ denotes a finite perturbation of the modular energy $K$ with respect to its vacuum expectation value. 

  In this paper we will show that these fluctuations in the modular Hamiltonian are due to the fluctuations in the spacetime geometry near light fronts that take the form of gravitational shockwaves.  The fluctuating shockwave geometries around the vacuum are a direct consequences of uncertainty relations between light ray operators, corresponding to the shifts $\delta u$ and $\delta v$, according to Eq.~(\ref{eq:uncertainty}).  
We find that the fluctuating shockwaves that are a consequence of the light-ray uncertainty relation  imply $\langle \Delta K^2 \rangle = \langle K \rangle =  {Area\over 4 G}$ for empty space.  We take this as evidence for fundamental uncertainties in the light ray operators as applied to the vacuum state. 

The outline of this paper is as follows.  In Sec.~\ref{sec:action} we review the shockwave action, long established in the literature, adapted for our purposes of studying spacetime behavior in a causal diamond.  Then, in Sec.~\ref{subsec:onshell} we show that, in planar coordinates, the shockwave action is precisely the modular Hamiltonian.  In Sec.~\ref{sec:commutators} we introduce quantum behavior by postulating commutation relations for the light front operators.  These quantum relations are closely related to those introduced by 't Hooft at black hole horizons.  Here we postulate these uncertainty relations describe light front operators at the horizons of causal diamonds.  This opens the way in Sec.~\ref{sec:mainresult} to compute expectation values of the modular Hamiltonian from the uncertainty in the light front operators, from which we obtain the results Eqs.~(\ref{eq:DeltaK}).  Finally, in Sec.~\ref{sec:spherical} we generalize our results to closed causal diamonds with spherical entangling surfaces, before concluding.
\section{Effective Action for Shockwave Geometries}
\label{sec:action}

	\begin{figure}[t]
\begin{center}
\includegraphics[scale=0.60]{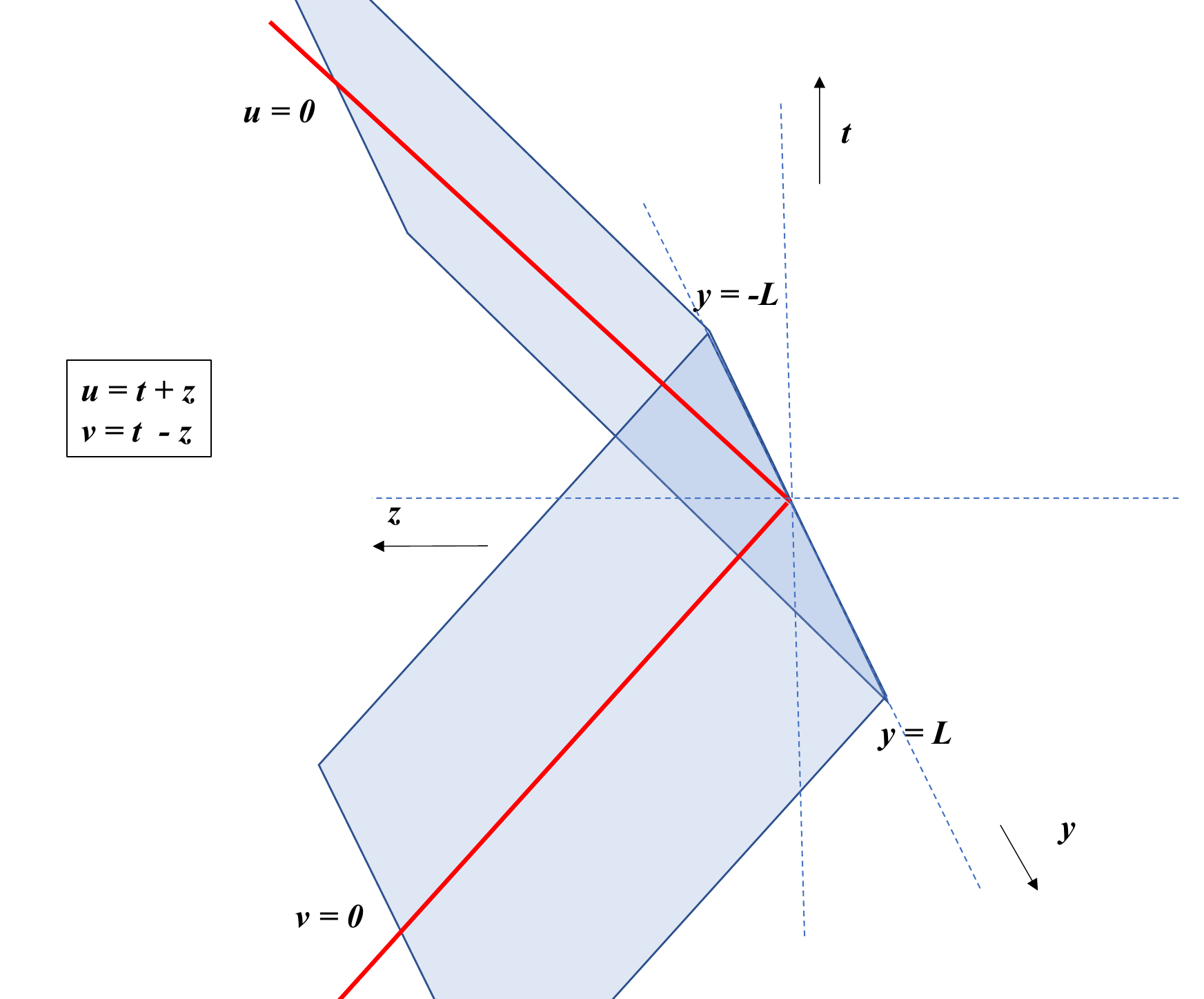}
\caption{ A planar lightsheet consisting of a family of light trajectories.  The directions longitudinal to the light sheet consist of the light cone coordinates, $u,~v$, while the transverse directions are labeled with $y$.}
\label{fig:planar-lightsheet}
\end{center}
\end{figure}

We are interested in the behavior of spacetime at light fronts, as shown in Fig.~\ref{fig:planar-lightsheet}.  We consider a family of light trajectories which first are traveling to the right and then are reflected back at $x=0$ at time $t=0$  to start moving to the left. 
In terms of the light cone coordinates $u=t+x$ and $v=t-x$, this means we are looking at trajectories that travel first from $u=-\infty$ to $u=0$ along $v=0$ and then from $v=0$ to $v=\infty$ along $u=0$.

Our goal is to determine the fluctuations in these trajectories due to quantum gravity effects. The fluctuations in the location of the trajectory due to a metric perturbations can be expressed as a change $\delta v$ in the value of the light cone coordinate for the lower trajectory, and a change $\delta u$ in upper trajectory. These coordinate shifts can be expressed as integrals of the metric perturbations $h_{uu}$ and $h_{vv}$ as follows
\beq
\delta v(u,y) = \int_{-\infty}^{u}\!\! \,du' \,h_{uu}(u',y), \qquad\quad \mbox{and} \qquad \quad \delta u(v,y) = \int_{-\infty}^{v}\!\! \,dv' \,h_{vv}(v',y). 
\eeq
We will assume that these metric perturbations are induced by fluctuations in the stress energy tensor $T_{uu}$ and take the form of a shock wave geometry, as given in Eq.~(\ref{eq:shockwavemetric}).
Note that $f(y,y')$ is only dimensionless in four dimensions.  A particularly clear example is the shockwave due to a fast particle moving  in the $v$-direction at transversal position $y''$. Its energy-momentum density is given by
\begin{equation}
	T_{uu}(u,y') = p_u \delta^{(d-2)}(y',y'')\delta(u-u_0)
\end{equation}
and produces shockwave geometry described by the Aichelburg-Sexl metric 
\begin{equation}
	ds^2 = -dudv + \ell_p^{d-2} p_u f(y,y') \delta(u-u_0) du^2+ dy^2.
\end{equation}
Geometrically a shockwave corresponds to two parts of flat Minkowski space 
glued together on the location of the shockwave after a coordinate shift $v\to v+\delta v$, where in the case of a single energetic particle
\begin{equation}
	\delta v(y) = \ell_p^{d-2} p_u f(y,y'').
\end{equation}
Similar and completely analogous equations hold for the shockwave geometry caused by a fast particle moving in the $u$-direction. 

We will be interested in general metric fluctuations that take the form of a shockwave geometry caused by fluctuations in the $T_{uu}$ and $T_{vv}$ stress energy tensor components. These geometries can be parametrized by functions $X^v(u,v,y)$ and $X^u(u,v,y)$ that incorporate the coordinate shifts due to the shockwave. For instance, $X^v$ combines the shift $\delta v(u,y)$ in the coordinate $v$ together with the (shifted) $v$-coordinate via
\be 
X^v(u,v, y) = v+\delta v(u,y),
\ee
Similarly we can introduce variables $X^u(u,v,y)$ describing the shock waves in the upper light trajectories. For a fast particle moving along fixed $(v,y)$ 
\be 
X^u(u,v, y) = u +\delta u(v,y),
\ee
a shift induced on the upper half of the causal diamond.  The general shockwave geometries can be conveniently written in a gauge in which one allows mixed transversal-longitudinal metric components\footnote{This form of the metric is derived from the one in terms of $h_{uu}$ and $h_{vv}$ by shifting the $u$ and $v$ coordinates by $\delta u$ and $\delta v$.} 
\begin{equation}
	ds^2 = -dudv + \nabla_y X^v du dy + \nabla_y X^u dv dy + dy^2. 
\end{equation}

\begin{figure}[t]
\includegraphics[scale=0.3]{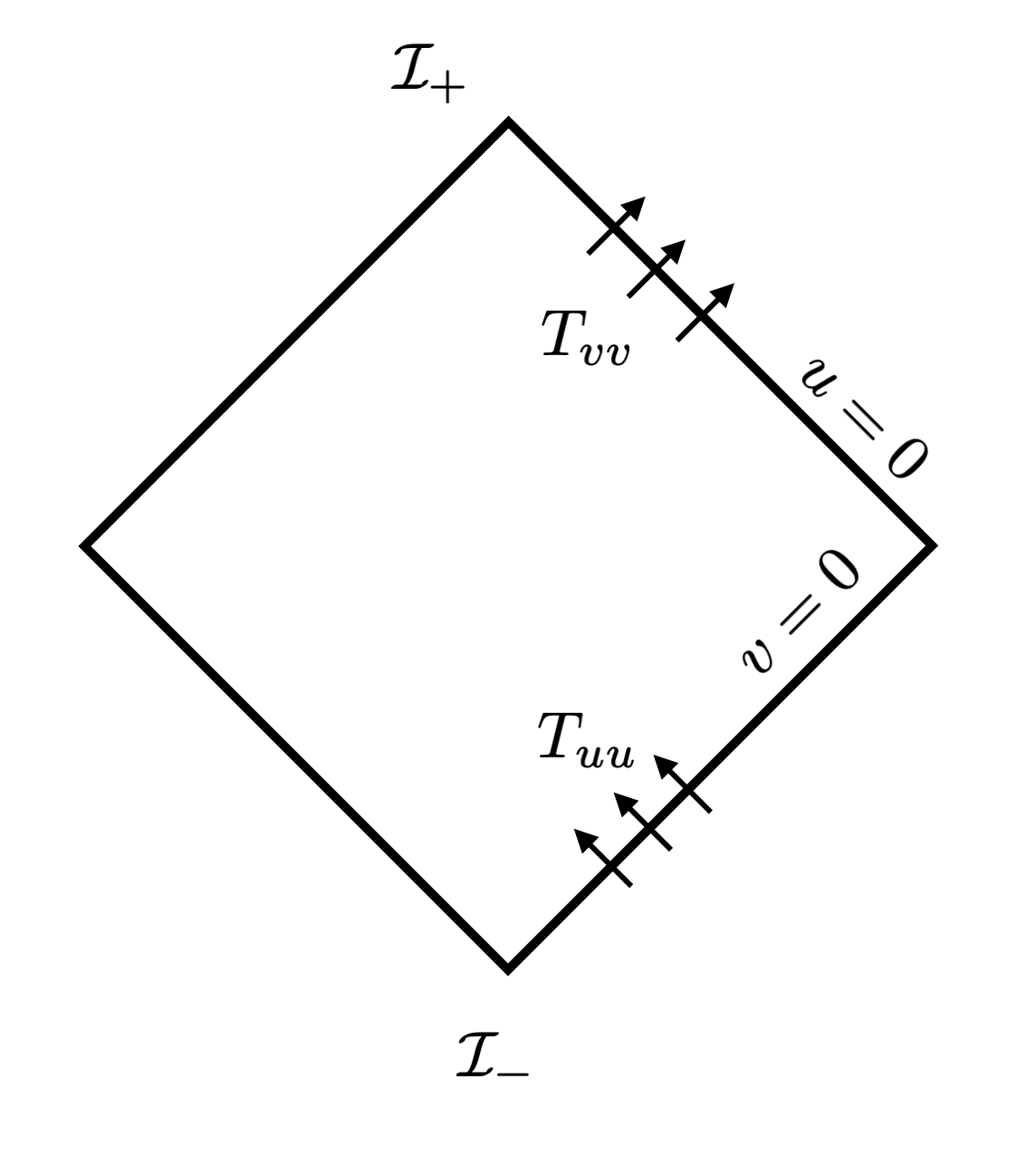}
\vspace{-0.2in}
\caption{We consider shockwave geometries of the type shown here, where vacuum fluctuations $T_{uu}(u,y)$ and $T_{vv}(v,y)$ induce shifts in the light cone coordinates $\delta v$ and $\delta u$ on the lower and upper half of the causal diamond, respectively.}
\label{fig:shockwave}
\end{figure}

The equations of motion for the variables $X^u$ and $X^v$ can be derived directly from the Einstein equations. Alternatively, one can derive an effective action for $X^u$ and $X^v$ by inserting the metric Ansatz for the shockwave geometry into the Einstein-Hilbert action. In this way one finds
\beq
\label{shockwave-action}
I = \int d^{d-2} y \left\lbrack -{1\over \ell_p^{d-2}}\int d\tau  \  X^u \Delta_y {dX^v\over d\tau} + \int\! d\tau\, \Bigl(X^u T_{u\tau} +X^vT_{v\tau }\Bigr)\,\right\rbrack 
\eeq
where 
$$
T_{u\tau}\equiv T_{uu} {du\over d\tau}+T_{uv} {dv\over d\tau}\qquad\mbox{and}\qquad T_{v\tau}\equiv T_{vu} {du\over d\tau}+T_{vv} {dv\over d\tau}.
$$
 This effective action does not live in the full $d$-dimensional spacetime, but on the $d-1$-dimensional boundary of the causal diamond.  Indeed, by inserting the metric Ansatz for the shockwave geometry one finds that the Einstein-Hilbert action becomes a total derivative, and hence is dimensionally reduced to one lower dimension. It is important to point out that the action is invariant under reparametrizations of the boundary time $\tau$. We will make use of this remark below.  In \cite{Verlinde:1991iu} a systematic derivation was given of the action (\ref{shockwave-action}) using a scaling argument appropriate for high energy scattering. We assume that these same scaling arguments apply to the present situation.  Note that there is an asymmetry in the time derivative acting on $X^u$ versus $X^v$, leading to a relative minus sign in the effective action depending on which variable the time derivative acts.  The physical reason, as shown in Fig.~\ref{fig:shockwave}, is that energy $T_{uu}$ is flowing into the causal diamond on the lower trajectory, but $T_{vv}$ is flowing out of the causal diamond on the upper trajectory.
 
 \subsection{The On-shell Action Equals the Modular Hamiltonian}
 \label{subsec:onshell}
 
One can easily verify that the action Eq.~(\ref{shockwave-action}) leads to the correct equations for the shockwaves. For this discussion it will be convenient to write the total action as a sum of three parts: one living on the past light trajectory, one part living on the future light trajectory, and a third term that is associated with the bifurcate horizon: 
\begin{equation}
	I = I_{lower} + I_{upper} + I_{horizon}.
\end{equation}
The horizon contribution comes from the asymmetry between $X^u$ and $X^v$ in the shockwave action Eq.~(\ref{shockwave-action}), where $X^v$ has a time derivative but $X^u$ does not.   An integration by parts to obtain the $X^u$ equation of motion generates $I_{horizon}$.  We now describe each of these contributions in turn.

For the unperturbed light trajectories, $v$ is constant on the lower part, while $u$ is constant on the upper part. First let us concentrate on the lower part of the causal diamond. We can then choose $\tau = u$, so that the action becomes
\be
I_{lower} =\int d^{d-2} y \left\lbrack \, - {1\over \ell_p^{d-2}} \int_{-\infty}^0\!\! du \,X^u \Delta_y {dX^v\over du} +\int_{-\infty}^0\!\! du \,X^u T_{uu} \,\right\rbrack. 
\ee
Here we have assumed that the stress energy tensor is effectively traceless, which in the scaling regime appropriate for shockwave means that $T_{uv}=0$.  One easily verifies that by varying $X^u$ in the action $I_{lower}$ one reproduces the correct shockwave equation for $X^v$
\be
\Delta_y {dX^v\over du} = \ell_p^{d-2} T_{uu}. 
\label{eq:EOM}
\ee
 The other equation of motion obtained by varying $X^v$ is also satisfied, since on this lower trajectory ${dX^u/du}$ is a constant.  When the equations of motion are satisfied one finds that the integrand of the action integral, and hence the action $I_{lower}$ itself, vanishes on-shell. 
 Similarly we define an action for the upper trajectory by interchanging the role of the $u$- and $v$-coordinates and replacing $X^u$ by $X^v$ and vice versa. This gives
\be
\label{upper-action}
I_{upper} = \int d^{d-2} y \left\lbrack \, {1\over \ell_p^{d-2}}\int_{0}^\infty\!\! dv \,X^v \Delta_y {dX^u\over dv} +\int_{0}^\infty\!\! dv \,X^v T_{vv}\, \right\rbrack 
\ee
This action leads, similarly as for the lower part,  to the correct equations of motion and vanishes on-shell, hence $I_{upper}=0$ on-shell. 
However, we have neglected a surface term in integrating by parts to obtain Eq.~\ref{upper-action} from Eq.~\ref{shockwave-action}, which takes the form: 
\begin{equation}
I_{horizon} = -{1\over \ell_p^{d-2}}\int \! d^{d-2}y \, X^u(y)\Delta_y  X^v(y), 
 \label{horizon-action}
\end{equation}
where we have denoted 
$$
X^v(y) \equiv X^v(0,0,y)\qquad \mbox{and} \qquad X^u(y) \equiv X^u(0,0,y).
$$

We will now show that the on-shell action can also be identified with the modular Hamiltonian $K$. The argument is as follows.  We will choose a new gauge where, instead of allowing $X^v$ and $X^u$ to fluctuate, we fix $X^v= 0$ at $u=-\infty$ for the lower trajectory in the absence of quantum fluctuations, and likewise $X^u =0$ for $v=\infty$ for the upper trajectory. 
This means that on shell the first term in the action $I$ vanishes. Hence, the total on shell action in that case just becomes
\begin{equation}
\label{onshell-action}
	I_{on-shell} = \int d^{d-2} y\left \lbrack \,\int_{-\infty}^0\!\! du \,X^u T_{uu}+\,\int_{0}^\infty\!\! dv \,X^v T_{vv}\,\right\rbrack  \equiv K
\end{equation}
The fields $X^u$ and $X^v$ can in fact be identified with the components of the Killing vector associated with the Killing horizon. 

Combining these two results leads then to an expression for the modular Hamiltonian $K$ in terms of the fluctuating shockwave variables $X^u$ and $X^v$ 
\begin{equation}
\label{Kshock}
	K= {1\over \ell_p^{d-2}}\int \! d^{d-2}y \, \nabla_y X^u(y)\nabla_y  X^v(y) 
\end{equation}
Here we rewrote the Laplacian as $\Delta_y=\nabla_y^2$ and performed a partial integration in the transversal plane. This result for the modular Hamiltonian may appear somewhat unexpected, but one should see it first of all as an on-shell relation that makes use of the Einstein equations to rewrite the stress energy in terms of the metric variables. 

In the following section we will also use it as an operator identity. In particular we will argue that the left and right hand side both have the same vacuum expectation value and also exhibit the same fluctuations. A key ingredient in our derivation will be commutation relations proposed by 't Hooft in the context of black hole horizons.  Here we apply them to a bifurcate light sheet horizon.

\section{Quantum Mechanics of Shockwave Operators}
\label{sec:commutators}

We will first present a summary of the commutation relations at the bifurcate horizon.  Let us introduce the momentum densities $P_u(u,y)$ and $P_v(v,y)$ via 
\begin{equation}
	P_u(u,y) = \int_{-\infty}^u \!\! du\, T_{uu}(u,y) \qquad\quad \mbox{and} \qquad\quad  P_v(v,y) = \int_{-\infty}^v \!\! dv\, T_{vv}(v,y)
\end{equation}
As quantum operators $P_u$ and $P_v$ generate shifts in the lightcone coordinates $u$ and $v$. It is natural to identify these shifts with the fields $X^u$ and $X^v$.  Following this reasoning one arrives at the equal time canonical commutation relations
\begin{equation}
	\Bigl\lbrack P_u(y), X^u(y')\Bigr\rbrack = i\, \delta^{(d-2)}(y,y') \qquad\quad \mbox{and} 
	\qquad\quad  \Bigl\lbrack P_v(y), X^v(y')\Bigr\rbrack = i\, \delta^{(d-2)}(y,y'). 
	\end{equation}	
On shell the momentum density operators $P_u$ and $P_v$ are related to the shockwave variables $X^u$ and $X^v$ via the equations of motion, which in terms of these variables read
\begin{equation}
\label{momentum-relations}
	\Delta_y X^u = \ell_p^{d-2} P_v \qquad\quad \mbox{and} \qquad\quad \Delta_y X^v = \ell_p^{d-2} P_u.
\end{equation}
Hence, the commutation relations can be rewritten directly in terms of the variables $X^u$ and $X^v$ as 
\begin{equation}
\label{hooftcommutator}
	\Bigl\lbrack X^u(y), X^v(y')\Bigr\rbrack = i\, \ell_p^{d-2} f(y,y').
\end{equation}
These are the 't Hooft commutation relations. The derivation is somewhat heuristic, since it makes use of the identification of the generator of translations in the coordinates $u$ and $v$ with the canonical conjugate to the variables $X^u$ and $X^v$. 

The justification of this heuristic derivation comes from the effective shockwave action (\ref{shockwave-action}). This action makes manifest that as operators $X^u$ and $X^v$ are non-commuting, since they both appear in the first term involving the time-derivative.  In fact, the relations Eq.~(\ref{momentum-relations}) directly identify the momentum density variables $P_u$ and $P_v$ with the canonical momenta conjugate to $X^u$ and $X^v$. In other words, by applying the standard canonical quantization rules to the effective action Eq.~(\ref{shockwave-action}) one precisely obtains the 't Hooft commutation relations.

Note that by the usual rules of quantum mechanics these commutation relations imply that the variables $X^u$ and $X^v$ have quantum mechanical uncertainties $\Delta X^u$ and $\Delta X^v$ that obey the uncertainty relations, which are directly related to the commutation relation via
\begin{equation}
	\Delta X^u(y) \Delta X^v(y') \geq  {1\over 2i}\Bigl\langle\Bigl \lbrack X^u(y),X^v(y')\Bigr\rbrack \Bigr \rangle 
\end{equation}	
The proof of this relation requires that $X^u(y)$ and $X^v(y)$ are hermitian operators: 
\beq
\label{Hermitian}
X^u(y) = \left(X^u(y)\right)^\dagger \qquad\quad\mbox{and}\qquad\quad X^v(y) = \left(X^v(y)\right)^\dagger. 
\eeq
One can note further that the uncertainty relations imply that the two point functions  $\langle X^u(y) X^u(y')\rangle$ and $\langle X^v(y) X^v(y')\rangle$ are generically non-vanishing. Their value, however, is not a priori determined and generally depends on the choice of state. 
In the following we are interested in verifying the calculation of the expectation value and fluctuations of the modular Hamiltonian. For this purpose we will make use of the Euclidean path integral over the shockwave variables $X^u$ and $X^v$. As we will see, this will lead to expressions that may appear counter-intuitive from the viewpoint of the standard Lorentzian quantization. In particular, we will see that in the Euclidean path integral the only non-zero two point function is $\langle X^u(y) X^v(y')\rangle$. Indeed, non-zero two point functions of the type $\langle X^u X^u\rangle$ or $\langle X^v X^v\rangle$ would break Lorentz invariance, and hence are only possible in situations with a preferred Lorentz frame.

\subsection{ From Lorentzian to Euclidean quantization}

The basic variables $X^u$ and $X^v$ can be continued to Euclidean space as follows. First we introduce 
\beq
X^u(y) = Z(y)+T(y)\qquad\qquad X^v(y) = -Z(y)+ T(y).
\eeq
We now replace $T\to iT_E$, so that
\beq
X^u_{{}_E}(y) = Z(y)+iT_{{}_E}(y)\qquad\qquad X^v_{{}_E}(y) = -Z(y)+iT_{{}_E}(y).
\eeq
We thus find that, instead of the relations (\ref{Hermitian}), the Euclidean operators now satisfy the following hermiticity property 
\beq
X^u_{{}_E}(y) = -\left(X^v_{{}_E}(y)\right)^\dagger.  
\eeq
The minus sign is just a choice of convention and will not have major implications. 
Euclidean quantization then proceeds by imposing canonical commutation relations
\beq
\label{euclideancommutator}
\Bigl\lbrack X^u_{{}_E}(y), X^v_{{}_E}(y')\Bigr\rbrack = l_p^2 f(y,y'),
\eeq
without the usual factor $i$. Hence, instead of being like coordinate and momentum variable, we now note that $X^u$ and $X^v$ behave as creation and annihilation operators. We can now persue this analogy and introduce a vacuum state that is annihilated by $X^v$, and its conjugate state that is annihilated by $X^u$. 
\beq
X_{{}_E}^v(y)|0\rangle =0\quad\qquad\mbox{and}
\quad \qquad \langle 0| X_{{}_E}^u(y) =0.
\eeq
Note this choice of initial and final states is consistent with the boundary conditions imposed on the fields $X^u(y)$ and $X^v(y')$ (see discussion just above Eq.~(\ref{onshell-action})). 

The goal is now to compute the two point functions. First of all, one immediately sees that in the Euclidean quantization there are no two-point functions of the type $\langle X^u X^u\rangle$ of $\langle X^vX^v\rangle$. The only non-vanishing two point function is 
\beq
 \langle 0| X_{{}_E}^v(y) X_{{}_E}^u(y')|0\rangle = l_p^2 f(y,y').
\eeq
which is directly derived from the commutator Eq.~(\ref{euclideancommutator}).  In the remainder of this paper we will be working in Euclidean signature, and hence we will drop the subscript $E$, leaving it implicit in our notation.
We will now show that the same result follows from the Euclidean path integral. 

\subsection{The Euclidean two point function}

 Since the effective action is quadratic, the path integral can be computed by applying Wick's theorem so that all correlators become expressed in terms of two point functions.  The two point function of $X^u$ and $X^v$  corresponds to the inverse of the kinetic operators,  and hence, in Euclidean space, obeys
\begin{equation}
	{1\over \ell_p^{d-2}} \Delta_y {d\over d\tau} \Bigl\langle X^u(\tau, y)X^v(\tau',y')\Bigr\rangle = \delta (\tau-\tau')\delta^{(d-2)}(y,y').
	\label{eq:XuXvcorrelator}	
\end{equation}
Note that in Euclidean space this two point function must be real.
Eq.~(\ref{eq:XuXvcorrelator}) is easily solved and leads to
\begin{equation}
	\Bigl\langle X^u(\tau, y)X^v(\tau',y')\Bigr\rangle = \ell_p^{d-2} \theta(\tau - \tau')f(y,y').
\end{equation}

Our goal is to show that the fluctuations in the modular Hamiltonian have their origin in tiny shockwaves in the spacetime geometry caused by quantum gravity effects. In particular, we want to derive the correct magnitude of both the expectation value $\langle K\rangle$ as well as the fluctuations $\langle \Delta K^2\rangle$. For this purpose it is sufficient to know the two-point functions of the coordinate shifts $X^u$ and $X^v$ at the bifurcate horizon. Hence we may ignore the time-dependence, and only consider the two point correlator as function of the transversal coordinates.  
Concretely this means we will take the limit where $u(\tau)\to L$ from below and $v(\tau')\to -L$ from above, so that $\theta(\tau-\tau') =1$. 
In this way we find 
\begin{equation}
\label{shockcorrelator}
	\Bigl\langle X^u( y)X^v(y')\Bigr\rangle = \,\ell_p^{d-2} f(y,y').
\end{equation}
We will use this result below to compute the expectation value and fluctuations of the modular energy. However, to obtain finite results we will need to introduce an extra ingredient. Namely, we will have to impose a short distance cut off on the allowed shockwaves geometries. Our analysis clearly breaks down for shockwaves whose transversal wavelength becomes shorter than the Planck scale. Indeed, it is natural to assume that one has to introduce a cut off on the allowed transversal wavelengths at or close to the Planck scale. Below we will indeed find that this cut off is necessary to find agreement with the known values for the modular energy fluctuations.

\section{Modular Energy Fluctuations from Shockwaves} 
\label{sec:mainresult}

We have now all the ingredients to compute the fluctuations 
of the modular Hamiltonian using the expression (\ref{Kshock}) and the results of the previous section. First we will consider the vacuum expectation value $\langle K\rangle$. In fact, for this computation we need to set the short distance cut off at its most natural value, which then turns out to precisely give the expected result. Once the correct value of the cut-off is known, we will proceed to compute the fluctuations. 

\begin{figure}[t]
\includegraphics[scale=0.3]{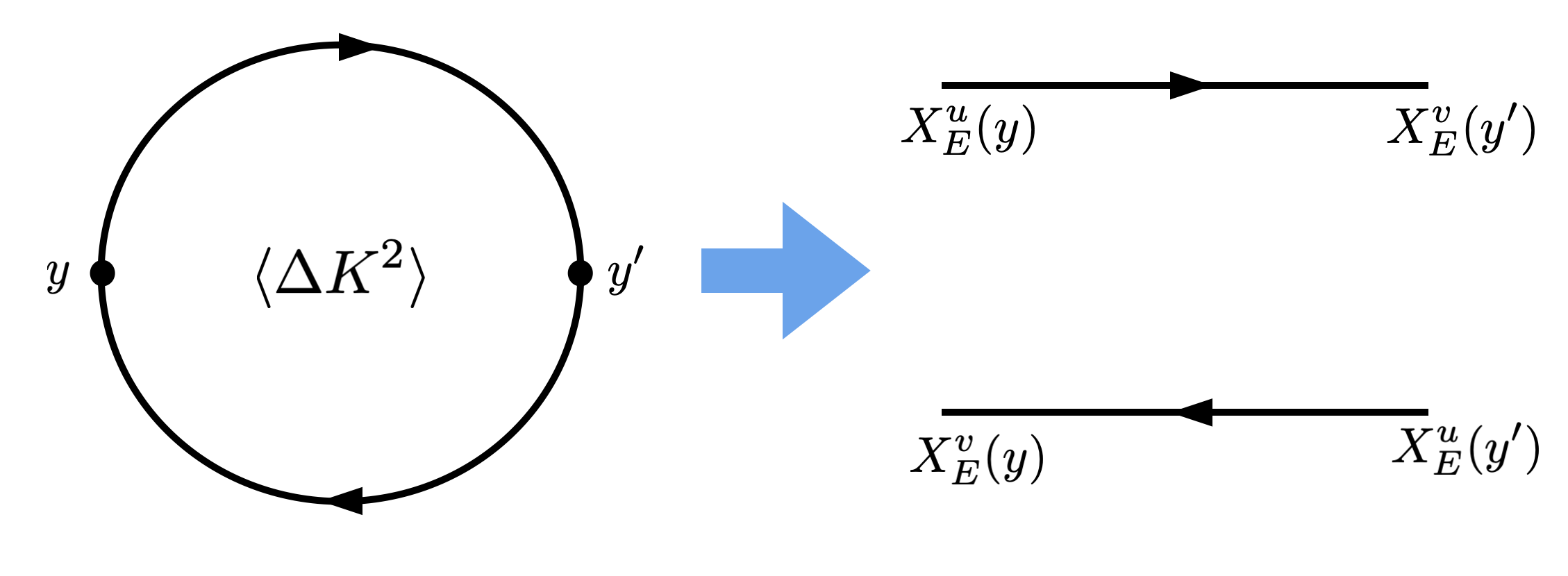}
\label{fig:deltaksq}
\caption{A pictorial representation of Eq.~(\ref{eq:deltaksqdecomp}), where the vacuum fluctuation of the modular Hamiltonian decomposes into the product of the two-point functions of the light ray operators $X^u,~X^v$ in Euclidean signature (denoted here by subscript $E$).}
\end{figure}

Combining the expression (\ref{Kshock}) with (\ref{shockcorrelator}) leads to the result 
\begin{equation}
	\bigl \langle K\bigr \rangle = {1\over \ell_p^{d-2}} \int d^{d-2} y \lim_{y'\to y} \nabla_y \nabla_{y'} \bigl \langle X^u(y) X^v(y')\bigr\rangle = \int d^{d-2} y \lim_{y'\to y} \nabla_y \nabla_{y'} f(y,y')
\end{equation}
The integrand in the right hand side formally diverges. But we will impose the most natural cut-off by assuming that the points $y$ and $y'$ need to be separated by at least one Planck length. This implies that 
\begin{equation}
	\lim_{y'\to y} \nabla_y \nabla_{y'} f(y,y') \, \sim \, {1\over \ell_p^{d-2}}.
\end{equation}
We will now choose the overall constant so that we obtain the expected result
\begin{equation}
	\bigl \langle K\bigr \rangle ={Area\over 4G}.
\end{equation}
Note that we have not really derived the constant of proportionality. But this is the only point were we can use the freedom to fix this constant. So to check that our proposed identification between the modular Hamiltonian and the fluctuating shock wave geometries is correct, let us now compute the variance $\langle \Delta K^2\rangle$ using the same method. In our calculation we will assume that the fluctuations are Gaussian, which means that all higher point correlation functions can be reduced via Wick's theorem to two-point functions. Since our calculation is being performed in the Euclidean setting, the only non-vanishing two point functions are between $X^u(y)$ and $X^v(y')$. 
By again combining the equations (\ref{Kshock}) and (\ref{shockcorrelator}) one obtains after a few steps
\begin{eqnarray}
\label{eq:deltaksqdecomp}
	\bigl \langle K{}^2\bigr \rangle - \bigl \langle K\bigr \rangle^2 & = & {1\over \ell_p^{2(d-2)}}\int d^{d-2}y \int d^{d-2}y' \ \Bigl  \langle \nabla_y X^u(y) \nabla_y X^v(y)\nabla_{y'} X^v(y') \nabla_{y'} X^u(y')\Bigr\rangle\nonumber\\
	&  & -{1\over \ell_p^{2(d-2)}}\int d^{d-2}y \int d^{d-2}y' \ \Bigl  \langle \nabla_y X^u(y) \nabla_y X^v(y)\Bigr\rangle\Bigl\langle\nabla_{y'} X^v(y') \nabla_{y'} X^u(y')\Bigr\rangle\nonumber
	\\{} & = &{1\over \ell_p^{2(d-2)}}\int d^{d-2}y \int d^{d-2}y' \ \nabla_y \nabla_{y'} \bigl \langle X^u(y) X^v(y')\bigr\rangle \nabla_y \nabla_{y'} \bigl \langle X^v(y) X^u(y')\bigr\rangle. 
	\end{eqnarray}
	In the last line we used that one of the two possible Wick contractions of the four-point function cancels against the product of the two-point functions. In particular this means that the Wick contractions between $X^u(y)$ and $X^v(y)$ at coincident points are being subtracted.  In other words, the only remaining Wick contractions are those between operators at different positions $y$ and $y'$. A pictorial representation of the decomposition in Eq.~(\ref{eq:deltaksqdecomp}) is shown in Fig.~\ref{fig:deltaksq}, where the vacuum fluctuation in the modular Hamiltonian becomes the product of two-point functions of the light-ray operators $X^u,~X^v$.   We can now insert the expression for the two-point functions in terms of the Green functions to obtain 	
	\begin{equation}
	\bigl \langle \Delta K^2\bigr \rangle 	= \int d^{d-2}y \int d^{d-2}y' \ \Bigl(\nabla_y \nabla_{y'} f(y,y')\Bigr)^2  \, =\, \int d^{d-2} y \lim_{y'\to y} \nabla_y \nabla_{y'} f(y,y').
	\end{equation}
	To obtain the second expression we performed a partial integration and used the fact that the Green function $f(y,y')$ obeys the identity (\ref{Green}). On the right hand side of this last equation we recognize the same expression that we obtained in the computation of $\langle K\rangle$.  By choosing the same natural value for the short distance cut off as before, we obtain the same result for the fluctuations in the modular Hamiltonian as its expectation value. Concretely, we find
\begin{equation}
	\bigl \langle \Delta K^2\bigr \rangle ={Area\over 4G}
\end{equation}
and thus reproduce the known answer for the modular energy fluctuations.  This is the main result of the paper.

We can now translate our result into a statement about the size of the quantum fluctuations in the spacetime geometry. 
Let us consider the product of  $\nabla_y X^u$ and  $\nabla_y X^v$ averaged of the transversal plane: 
\begin{equation}
	\bigl\lbrack \nabla_y X^u\nabla_yX^v\bigr\rbrack_{avg}\equiv {1\over Area}\int \! d^{d-2}y \,\nabla_y X^u(y)\nabla_y  X^v(y). 
\end{equation}
	In fact, the quantities $\nabla_y X^u$ and  $\nabla_y X^v$ can be identified with the fluctuations in the metric components $g_y{}^u$ and $g_y{}^v$:
\begin{equation}
	\delta g_y{}^u=\nabla_y X^u\qquad\quad \mbox{and}\qquad\quad \delta g_y{}^u=\nabla_y X^u.
\end{equation}

Our main result can thus be reformulated as a statement about the size of the fluctuations in these metric components
\begin{equation}
\Bigl\langle 	\bigl(\delta g_y{}^u \delta g_y{}^v\bigr)^2
	\Bigr\rangle 	\sim  \Bigl\langle 	\bigl\lbrack \nabla_y X^u\nabla_yX^v\bigr\rbrack_{avg}^2
	\Bigr\rangle \sim \left ({\ell_p\over L}\right)^{d-2}
	\end{equation}
	Note that in $d=4$ we thus find that $\sqrt{\langle (\delta g^2)^2\rangle}\sim L/\ell_p$. This result was anticipated in previous work, and appears to coincide with the intuitive answer that would result from a random walk picture, in which Planckian fluctuations accumulate along the light trajectory over a distance $L$, as reviewed in Ref.~\cite{snowmass} 
	in a general number of dimensions.  This point of view becomes more manifest in a spherically symmetric set-up, in which the various contributing modes can be labeled by discrete quantum numbers. In the remainder of this paper we will describe this spherically symmetric perspective in more detail. 
	
	\begin{figure}[t]
\includegraphics[scale=0.4]{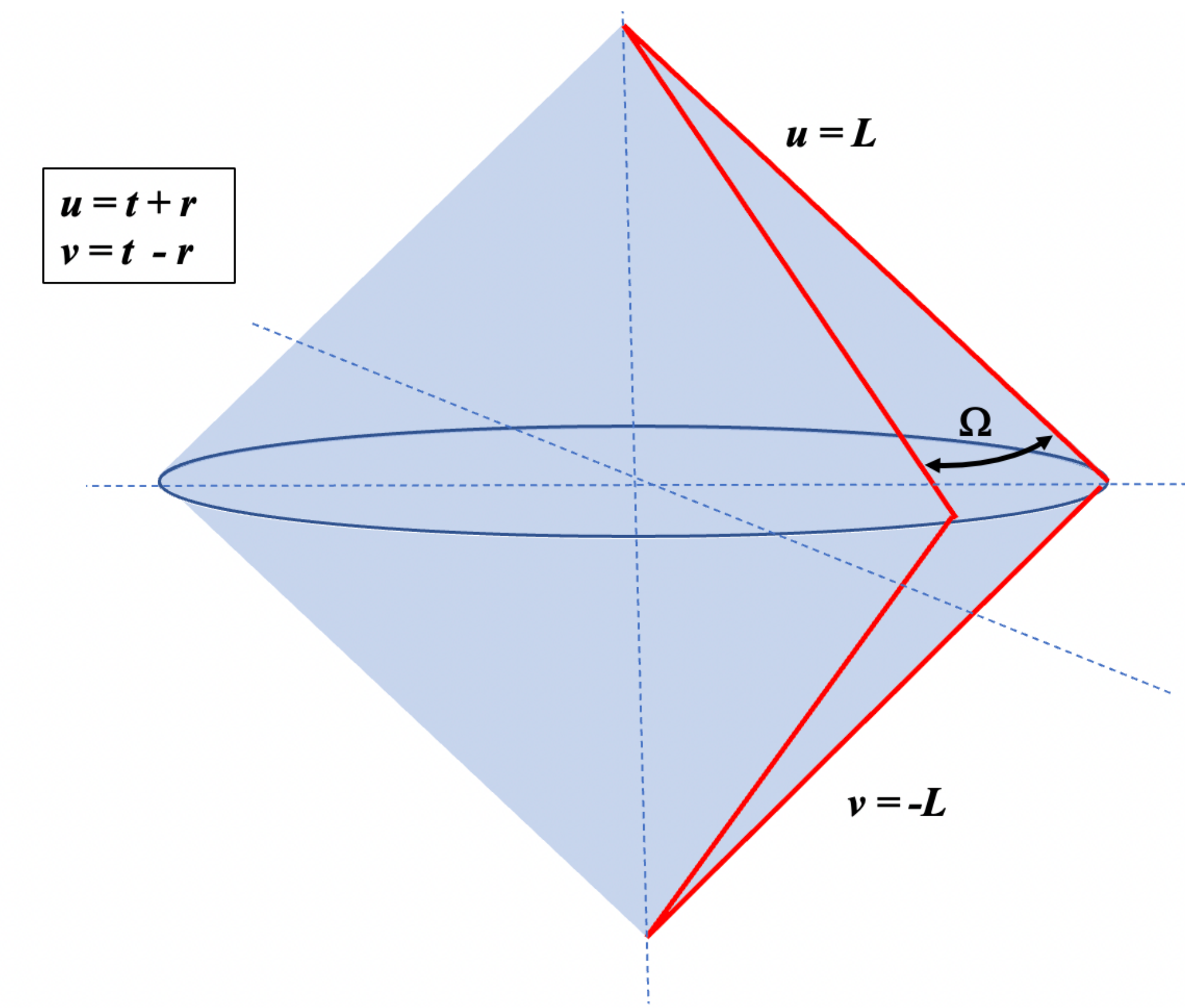} \vspace{-0.1in}
\label{fig:causaldiamond}
\caption{
A causal diamond with a spherical entangling surface of size $L$. The red lines describe two radially outward traveling light rays separated by an angle $\Omega$ that are reflected back at the entangling surface.  
}
\end{figure}


	\section{Causal Diamonds and Spherical Shockwaves}
	\label{sec:spherical}

We now consider a spherically symmetric causal diamond in which we place the light source and detector in the origin, and consider a family of light trajectories starting at $t=-L$ traveling in all directions before being reflected by to the origin at time $t$ by a family of mirrors located on a sphere of radius $L$. In this setup it is natural to work in spherical coordinates $(t,r,\Omega)$. The family of light rays can be depicted as traveling along the boundary of a causal diamond, which locally looks like a Rindler horizon.  It is worth commenting, as shown in Fig.~\ref{fig:causaldiamond}, that the difference in the clock of a Rindler observer Planckian separated from the horizon at the bottom of the causal dimaond, is given by the geometric mean of the Planck length with the size of the causal diamond.  The bifurcate horizon is now located at $t=0$ and corresponds to the sphere with radius $L$ centered around the origin.   

The calculations for the spherical configuration are very similar to the case with the flat Rindler horizon. We will therefore mainly focus on those aspects which are different. First of all, the boundary of the causal diamond strictly speaking does not correspond to a horizon, since it is not left invariant by a boost-like Killing vector. Instead, as is well described in the literature, it is invariant under a conformal Killing vector that locally near the boundary looks like a boost. We will therefore still refer to the boundary of the diamond as the horizon. 
The advantage of using a spherical set up is that the area of the bifurcate horizon is finite and given by the area of a $d-2$-sphere of radius $L$. Since our calculations will eventually reduce to the bifurcate horizon, we may consider fields $X^u(\Omega)$ and $X^v(\Omega)$ as functions of the angular coordinates $\Omega$: the transversal coordinates $y$ that were used in the previous sections may roughly be thought of as $y\sim L\Omega$. In particular, one finds that the role of the transversal Laplacian and its associated Green function are now taken over by 
\begin{equation}
	\Delta_y \quad \rightarrow \quad {1\over L^2}\, \bigl(\Delta_{\Omega}- R_\Omega\bigr)\qquad \mbox{and}\qquad f(y,y') \quad \rightarrow \quad {1\over L^{d-4}}\, f(\Omega,\Omega') 
\end{equation}
where $\Delta_\Omega$ and $R_\Omega$ are the scalar Laplacian and the Riemann curvature of the unit $d-2$-sphere: in general dimension $d$ one has $R_\Omega = (d-2)(d-3)/2$, 
while for $d=4$ the result is simply  $R_\Omega =1$.  The angular Green function $f(\Omega,\Omega')$ obeys
\begin{equation}
	(\Delta_\Omega -R_\Omega) f(\Omega,\Omega') =\delta^{(d-2)}(\Omega,\Omega').
\end{equation}
When we make these substitutions it follows that the Lagrangian description of the shockwave variables $X^u(\Omega)$ and $X^v(\Omega)$ only depends on a particular combination of the UV scale $\ell_p$ and IR scale $L$, namely 
\begin{equation}
 \tilde{\ell}_p^2 \, \equiv \, \frac{\ell_p^{d-2}}{L^{d-4}}.
\end{equation}
In particular, one finds the two point function of the coordinate shifts for the spherical case is given by 
\begin{equation}
	\Bigl \langle X^u(\Omega) X^v(\Omega')\Bigr\rangle  =  \tilde{\ell}_p^2 f(\Omega ,\Omega'). 
\end{equation}
Let us introduce the momentum surface density operator $P_u(\Omega)$ defined by
\begin{equation}
 P_v(
\Omega) = 	L^{d-2} \int_0
^L\!\! dx^v \, T_{vv}(v,\Omega \qquad\mbox{and}\qquad P_u(
\Omega) = 	L^{d-2} \int_0
^L\!\! dx^u \, T_{uu}(u,\Omega).
\end{equation}
The factor $L^{d-2}$ is included so that $P_u(\Omega)$ represent the momentum density per unit solid angle.  The equations of motion for the shockwave geometries can then be expressed as
\begin{equation}
	 \bigl(\Delta_{\Omega}-R_\Omega\bigr)X^u(\Omega) = \tilde{\ell}_p^2 \, P_v(\Omega)\qquad\mbox{and}\qquad  \bigl(\Delta_{\Omega}-R_\Omega\bigr)X^v(\Omega) = \tilde{\ell}_p^2 P_u(\Omega).
\end{equation}
These equations should again be seen as operator identities. As operators the momenta $P_u$ and $P_v$ are, just as for the flat case, canonically conjugate to $X^u$ and $X^v$.

\subsection{Modular Energy Fluctuations for Spherical Causal Diamonds}

We now discuss the computation of the modular energy fluctuations for causal diamonds. Following the same reasoning as for the flat horizon, one can argue that the modular Hamiltonian $K$ associated with the causal diamond may be expressed as
\begin{equation}
	K = -{1\over \tilde{\ell}_p^2}\int \!d^{d-2}\Omega\, X^u(\Omega) (\Delta_\Omega-R_{\Omega}) X^v(\Omega).
\end{equation}
To calculate the expectation value and fluctuations of $K$ we again need to introduce a Planckian cut-off. This is conveniently implemented by making use of the mode expansion of the variables in terms of spherical harmonics.

We will first describe this mode expansion and its consequences in $d=4$, and discuss the generalization to other dimensions afterwards. 
On the sphere we can write out the fields $X^u$ and $X^v$ as 
\begin{equation}
	X^u(\Omega) = \sum_{l,m} X^u_{lm}Y_{lm}(\Omega)\qquad \mbox{and}\qquad 	X^v(\Omega) = \sum_{l,m} X^v_{lm}Y_{lm}(\Omega),
\end{equation}
where $m$ takes the usual range from $-l$ to $l$. The modular Hamiltonian 
can now be written as a sum over the different modes as 
\begin{equation}
	K = \sum_{l,m} (l^2+l+1) X^u_{lm} X^*{}^{v}{}_{lm}.
	\end{equation}
Here we used the fact that the spherical harmonics $Y_{lm}$ are eigenstates of the modified laplacian $\Delta_{\Omega}-R_{\Omega}$ with eigenvalue $-(l^2+l+1)$.

The short distance cut-off is imposed by requiring that $l$ only takes values up to a certain maximam $l_{max}$. This leads for instance to a regulated Green function 
\begin{equation}
	f_{reg}(\Omega,\Omega') = \sum_{l=0}^{l_{max}}\sum_{m=-l
	}^l {Y_{lm}(\Omega) Y^*_{lm}(\Omega')\over l^2+l+1}.
\end{equation}
From this one can directly read off that the two-point function of the discrete variable $X^u_{lm}$ and $X^v_{lm}$ is given by
\begin{equation}
	\Bigl \langle X^u_{lm} X^{v}{}_{l'm'}\Bigr\rangle =\ell_p^2 \, {\delta_{l,l'}\delta_{m,-m'}\over l^2+l+1}.
\end{equation}
This leads to the following result for the expectation value of the modular Hamiltonian 
\begin{equation}
	\bigl \langle K\bigr \rangle = \sum_{l=1}^{l_{max}}\sum_{m=-l}^l (l^2+l+1) \Bigl \langle X^u_{lm} X^*{}^{v}_{lm}\Bigr\rangle = \sum _{l=0}^{l_{max}} (2l+1) = (l_{max}+1)^2.
\end{equation}
Let us for the moment leave the value of $l_{max}$ to be determined. 
Given a choice for $l_{max}$ it becomes possible to compute the fluctuations in the modular Hamiltonian in a similar way.  After some straightforwards steps one finds
\begin{equation}
	\bigl \langle \Delta K^2\bigr \rangle = \sum_{l=1}^{l_{max}}\sum_{m=-l}^l \sum_{l'=1}^{l_{max}}\sum_{m'=-l'}^{l'}(l^2+l+1)(l'^2+l'+1) \Bigl \langle X^u_{lm} X^*{}^{v}{}_{l'm'}\Bigr\rangle \langle X^u_{l'm'} X^*{}^{v}{}_{lm}\Bigr\rangle. 
\end{equation}
By inserting the result for the second two point function one easily verifies that the sum over $l'$ and $m'$ can be explicitly performed and leads to an identical result as for the expectation value of $K$. One gets
\begin{equation}
	\bigl \langle \Delta K^2\bigr \rangle = \sum_{l=1}^{l_{max}}\sum_{m=-l}^l (l^2+l+1) \Bigl \langle X^u_{lm} X^*{}^{v}_{lm}\Bigr\rangle = \sum _{l=0}^{l_{max}} (2l+1) = (l_{max}+1)^2.
\end{equation}
We thus find that both the expectation value as well as the fluctuations of $K$ depend in an identical way on the value of the mode cut off $l_{max}$. We can now choose the value of $l_{max}$ so that the expectation value of $K$ gives the expected result in terms of the area of the bifurcate horizon. As we have just shown, we then obtain the same result for the fluctuations, and thereby confirm the expected result obtained by other methods. 

These same methods apply to higher dimensional spacetimes and lead to identical conclusions. The only difference is that instead of the conventional spherical harmonics, one has to use their higher dimensional generalizations.

\subsection{Lorentzian Quantization and Uncertainty Relations}

In our calculations of the modular energy fluctuations we used a Euclidean quantization procedure, in which $X^u$ and $X^v$ are treated as hermitian conjugate variables. In Lorentzian signature, $X^u$ and $X^v$ are each Hermitian, and are only canonically conjugate variables. This means their commutation relations contain a factor $i$, as for the usual case of coordinates and momenta. In addition one can follow this analogy and conclude that in Lorentzian quantization the operators $X^u$ and $X^v$ must have quantum mechanical uncertainties $\Delta X^u$ and $\Delta X^v$ that obey uncertainty relations. In this subsection we will describe these uncertainty relations fo the case of the finite causal diamond. This is a particularly convenient situation, since the coordinate fields $X^u$ and $X^v$ have a mode expansion in terms of discrete set of modes $X^u_{lm}$ and $X^v_{l'm'}$ obeying canonical commutations relations. 
\begin{equation}
		\Bigl\lbrack  X^u_{lm}, X^{v}{}_{l'm'}\Bigr\rbrack =i\,\ell_p^2 \, {\delta_{l,l'}\delta_{m,-m'}\over l^2+l+1}.
\end{equation}
where we reinstated the factor of $i$. Applying the standard derivation of the Heisenberg uncertainty relations leads to
\begin{equation}
		\Delta X^u_{lm} \Delta X^{v}{}_{l'm'} \, \geq \,{\ell_p^2 \over 2} \, {\delta_{l,l'}\delta_{m,-m'}\over l^2+l+1}.
\end{equation}
When translated back to coordinate space one finds that the uncertainty relation implies that 
\begin{equation}
	\Delta X^u(\Omega) \Delta X^{v}{}(\Omega') \geq {l_p^2\over 2} f(\Omega,\Omega')
	\end{equation}
As mentioned before, these uncertainty relations also lead to inequalities to the conclusion that in Euclidean space two point functions type $\langle X^u(\Omega) X^u(\Omega')\rangle$ and $\langle X^v(\Omega) X^v(\Omega')\rangle$ are non-vanishing, and represent fluctuations that are associated with only the lower or only the upper trajectory. The study of the physical implications of these type of fluctuations will be left for further work. 

\subsection{Generalization to Other Dimensions}
The spherical harmonics $Y_l$ on a higher dimensional sphere are labeled by an integer $l$. Let us consider the unit $(d-2)$-sphere $S^{d-2}$ contained in ${\mathbf R}^{d-1}$. One can represent the harmonic functions on $S^{d-2}$ as restrictions of the set of solutions to the $d-1$-dimensional Laplace equation $\Delta p(x)=0$, where $p(x)\in {\bf P}_l$ is a polynomial of degree $l$. The eigenvalues of the spherical Laplacian are 
\begin{equation}
	\Delta_{S^{d-2}} Y_l = -l(l+d-3) Y_l.
\end{equation}
We denote the space of all independent spherical harmonics $Y_l$ by $H_l$. The number of independent spherical harmonics $Y_l\in H_l$ with a given value of $l$ equals
\begin{equation}
{\rm dim}\, H_l 	= \left( \begin{array}{c}\!l\!+\!d\!-\!2\\\!d \!-\! 2\! \end{array}\right)-\left( \begin{array}{c}\!l\!+\!d\!-\!4\!\\ \!d\! -\! 2\! \end{array}\right).
\end{equation}
This generalizes the familiar case for $d=4$. 

We can now use the space of spherical harmonics to put a mode cut off on the Green function and delta functions on the sphere in higher dimensions. The Green functions are again given by
\begin{equation}
f(\Omega,\Omega') = \sum_{l}\sum_{Y_l\in H_l} {Y_l(\Omega) Y_l(\Omega')\over l(l+d-3) + (d-2)(d-3)/2}.	
\end{equation}
where we added the contribution of the Ricci curvature to the eigenvalue of the laplacian. Similarly as as for  $d=4$ we put a mode cut off of $l$ so that $l\leq l_{max}$. The computation of the expectation value and fluctuation of $K$ proceeds identically as for $d=4$ and will not be repeated here. The answer is again given by the sum of the dimensions of all the representations with $l\leq l_{max}$. The result is
\begin{equation}
\bigl\langle  K\bigr \rangle =\bigl\langle \Delta K^2\bigr\rangle= \sum_{l=0}^{l_{max}} {\rm dim}\, H_l 	= \left( \begin{array}{c}\!l_{max}\!+\!d\!-\!2\!\\ \!d -2\! \end{array}\right)\!+\!\left( \begin{array}{c}\!l_{max}\!+\!d\!-\!3\!\\ \!\!d -2\!\! \end{array}\right)\sim {1\over (d\!-\!2)!} \ l_{max}^{d-2}. \\[1mm]
\end{equation}
\noindent
So, just like in four dimensions one finds that, in order to reproduce the expected result given by the area of the bifurcate horizon, one has to choose $ l_{max} \sim {L/ \ell_p}$. But once the value of $l_{max}$ is fixed to give the correct expectation value of $K$, the result for the fluctuations in $K$ automatically comes out correctly as well. 

\section{Conclusions}

We have shown that shockwave geometries give rise to fluctuations in the modular Hamiltonian, $\langle K \rangle = \langle \Delta K^2 \rangle$.  The shockwave geometries are generated by vacuum fluctuations of a size given by uncertainty relations in light ray operators, Eq.~(\ref{hooftcommutator}).  Since $\langle K \rangle = \langle \Delta K^2 \rangle$ has been by now well-established in many contexts, including for boundary-anchored diamonds in AdS/CFT \cite{perl,deBoer:2018mzv,Nakaguchi:2016zqi,VZ2}, and near light fronts in flat space \cite{BZ}, our result supports the idea that the commutator in Eq.~(\ref{hooftcommutator}) is the fundamental object governing the quantum mechanics of spacetime. 

In previous work, we argued that modular fluctuations could give rise to uncertainties in the location of light fronts that {\em accumulate} into the infrared \cite{VZ1,snowmass}, becoming observably large over the light crossing time of a causal diamond.  The fundamental uncertainty relation in Eq.~(\ref{hooftcommutator}) effectively acts as a noise term for the causal development of a region of spacetime, giving us a new tool to compute quantum uncertainties in position observables.  We leave such an application for future work.

\mysection{Acknowledgments}
We thank Tom Banks, Yanbei Chen, Temple He, Cynthia Keeler, Vincent Lee and  Allic Sivaramakrishnan for discussion on these directions. We are supported by the Heising- Simons Foundation ``Observational Signatures of Quantum Gravity" collaboration grant 2021-2817.  The work of KZ is also supported by a Simons Investigator award and the U.S. Department of Energy, Office of Science, Office of High Energy Physics, under Award No. DE-SC0011632.

\bibliography{QG}

\end{document}